\documentclass[3p,times,procedia]{elsarticle}
\usepackage{ecrc}
\usepackage[bookmarks=false]{hyperref}
    \hypersetup{colorlinks,
      linkcolor=blue,
      citecolor=blue,
      urlcolor=blue}

\usepackage[table]{xcolor}
\definecolor{maroon}{cmyk}{0,0.87,0.68,0.32}      
\firstpage{1}

\journalname{Procedia Computer Science}

\runauth{M. Spichkova}

\jid{procs}

\usepackage{amssymb}
\usepackage[figuresright]{rotating}
\usepackage{xcolor}
\usepackage{tabularx}
\usepackage{lscape}
\usepackage{longtable}
\usepackage{supertabular}
\usepackage{makecell}
\usepackage{graphicx}
\usepackage{float}
\usepackage{multirow}
\usepackage{tikz}
\usepackage{hyperref}
\usepackage{url}
\usepackage{placeins}
\usepackage{array}
\usepackage{booktabs}
\usepackage{siunitx}

\begin{document}
\begin{frontmatter}
\dochead{29th International Conference on Knowledge-Based and Intelligent Information \& Engineering Systems (KES 2025)}%

\title{Agile and Student-Centred Teaching of Agile/Scrum Concepts}
\author[a]{Maria Spichkova*} 
\address[a]{School of Computing Technologies, RMIT University, Melbourne, Australia}

\begin{abstract}
In this paper, we discuss our experience in designing and teaching a course on Software Engineering Project Management, where the focus is on Agile/Scrum development and Requirement Engineering activities. The course has undergone fundamental changes since 2020 to make the teaching approach more student-centred and flexible.
As many universities abandoned having face-to-face exams at the end of the semester, authentic assessments now play an even more important role than before. This makes assessment of students' work even more challenging, especially if we are dealing with large cohorts of students. The complexity is not only in dealing with diversity in the student cohorts when elaborating the assessment tasks, but also in being able to provide feedback and marks in a timely and fairly. 
We report our lessons learned, which might provide useful insights for teaching Agile/Scrum concepts to undergraduate and postgraduate students. We also analyse what course structure might be effective to support a blended learning approach, as well as what could be a reasonable structure of online assessments, to keep them both authentic and scalable for large cohorts of students. \\ 
\emph{Preprint. Accepted to the 29th International Conference on Knowledge-Based and Intelligent Information \& Engineering Systems (KES 2025). Final version to be published by Elsevier (In Press).} 
\end{abstract}

\begin{keyword}
SE education, student-centred teaching, software engineering, requirements engineering, Agile, Scrum
\end{keyword}
\cortext[cor1]{Corresponding author.}
\end{frontmatter}

\email{maria.spichkova@rmit.edu.au}

\section{Introduction}

   \begin{figure}[ht!]
  \centering  
 \includegraphics[width=0.45\linewidth]{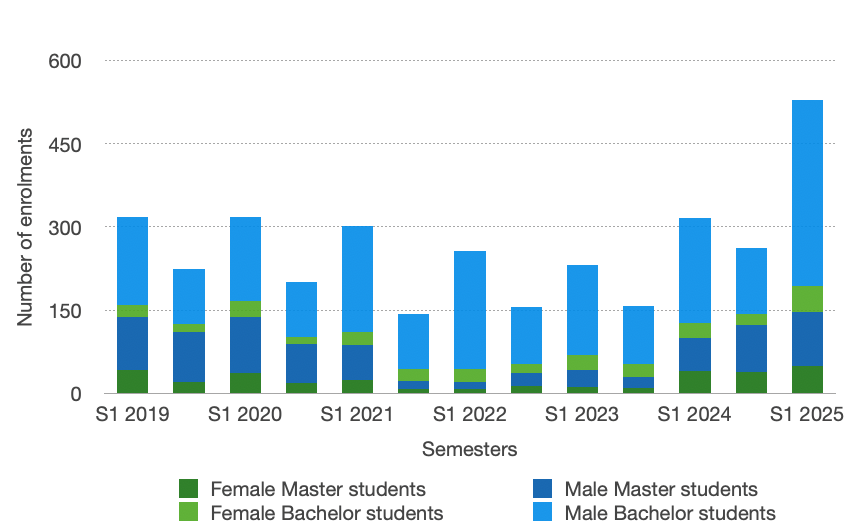}
    \caption{Enrolments statistics (S1 and S2 denote  Semesters 1 and 2 respectively)}
     \label{fig:enrolments}
 \end{figure}

Since 2020, the working style in the industry has changed to be either hybrid or fully online. This, on one side, changed the expectations of the graduates, who should be ready to collaborate online, and university education should prepare students for this work style. On the other side, many universities abandoned having face-to-face exams at the end of the semester, and this change requires even more careful assessment design. 
To better prepare students for their future careers, the assessments should be \emph{authentic}, i.e. they should replicate the tasks and performance standards typically faced by professionals in industry, see \cite{wiggins1990case,villarroel2018authentic}. 
Providing authentic assessments  as online tasks is
especially challenging when we are dealing with large cohorts of students, as we need to try to exclude plagiarism and contract cheating. 
Another challenge of authentic assessment is providing feedback from the tutors and marking results in a very short time frame. This is critical both (1)~within the semester when students are keen to receive timely and comprehensive feedback for improving their skills as well as for preparing for end-of-semester assessments, and (2)~at the very end of the semester when the students' grades have to be finalised for each course. On top of this, we also have budget/ workload constraints -- the course coordinators typically have a limited number of tutors/teaching assistants (if any), and tutors have a very limited capacity for marking tasks.

In this paper, we report on six years of teaching a course \emph{ISYS 1106 /1108/ 3474  Software Engineering Project Management}, which focuses on Agile/Scrum development, Software Engineering (SE) and Requirement Engineering (RE) activities. The course is a
mandatory course for Bachelor's and Master's students in the IT and Cybersecurity programs at RMIT University (Australia). It has been provided each semester, with the number of enrolments previously being around 150 to 300 students, while in Semester 1 2025, the number of enrolments became above 500, see Figure~\ref{fig:enrolments} for the enrolment statistics. 
 The course went through fundamental changes since 2020: in 2020-2021, it was provided in the online mode, in 2021-2022 in the hybrid mode, and since 2023 it is back to the face-to-face mode, with all lectorial classes available also as recordings.   
Based on our findings over 2019-2024, we present lessons learned as well as concrete recommendations on a student-centred approach for teaching SE courses with online assessments. 
In our reflection, we especially focused on the following research questions: 

\emph{\textbf{RQ1:}
 How to structure a course to support blended learning? To what extent in what setting a blended learning approach is effective for SE/IT courses?} Blended learning~\cite{dangwal2017blended,hrastinski2019we,rasheed2020challenges}, where the classroom face-to-face learning experiences are integrated with online experiences, became a compulsory component in many universities over the last years. This approach has many benefits as well as challenges to implement.  

\emph{\textbf{RQ2:} 
What could be a reasonable structure of online assessments, to keep them both authentic and scalable for large cohorts of students?} 
It's often considered that online assessments would lead to grade inflation and an unreasonable increase in the pass rate because of cheating. After the improvements we made to our course, we could
observe that the rate of students who passed the course is on a similar level what it was before the introduction of online-only assessments. For the Bachelor cohort, it is actually even a bit lower and is around 85\%, which we consider reasonable for this course. 
The mean grade of the Master cohort increased, however, this is rather related to the cohort changes: over the last semesters, the Master student cohort (where the percentage of international students was typically high) became smaller but more dedicated to learning.

\section{Course Setup: Focus on Project-based learning} 
\label{sec:course-structure}

In this section, we provide a brief overview of our course ISYS 1106 /1108/ 3474 on Software Engineering Project Management.   
 The course is dedicated to teaching project management concepts and methodologies, tools, as well as corresponding requirements engineering and quality assurance approaches. The focus of the course is on Agile/Scrum, but we also teach traditional Waterfall, Lean, RUP, and hybrid methodologies. 
 The expected Course Learning Outcomes include
  \begin{enumerate}
      \item 
      Illustrate a working knowledge of how to plan, execute and close projects to required standards;
       \item  
Use a range of tools to carry out and report on team projects;
 \item 
Use project management frameworks that ensure successful outcomes.
\item 
Analyse and discuss critical project management concepts, such as: Why Projects Fail; Project Governance and Methodologies; Software Development Life Cycles – From Waterfall to Agile; Software Engineering Fundamentals; Software Requirements Engineering; Planning and Scheduling; Risk and Issues Management; Quality Assurance; Change Management; Release Management; Service Delivery and Support; The Team Dynamic; Collaboration and Communication skills; Organisations, People and Culture.
\item 
Apply critical analysis, problem-solving, and team facilitation skills to software engineering project management processes using real-world scenarios.
  \end{enumerate}
The course is worth 12 credit points in the program curriculum, where the semester duration (teaching period) is 12 weeks. Thus, it is expected that students will invest approx. 120 hours of their work in this course. 
One of the complexities of teaching that course is the diversity of the cohorts:  they include undergraduate and postgraduate students from several programs, e.g., Computer Science, IT, SE and Cybersecurity. As illustrated in Figure~\ref{fig:enrolments}, the majority of the enrolments each semester are male Bachelor students, whereas the smallest cohort typically is female Master students. %
 
\begin{figure}[ht!]
  \centering  
  \includegraphics[width=0.65\linewidth]{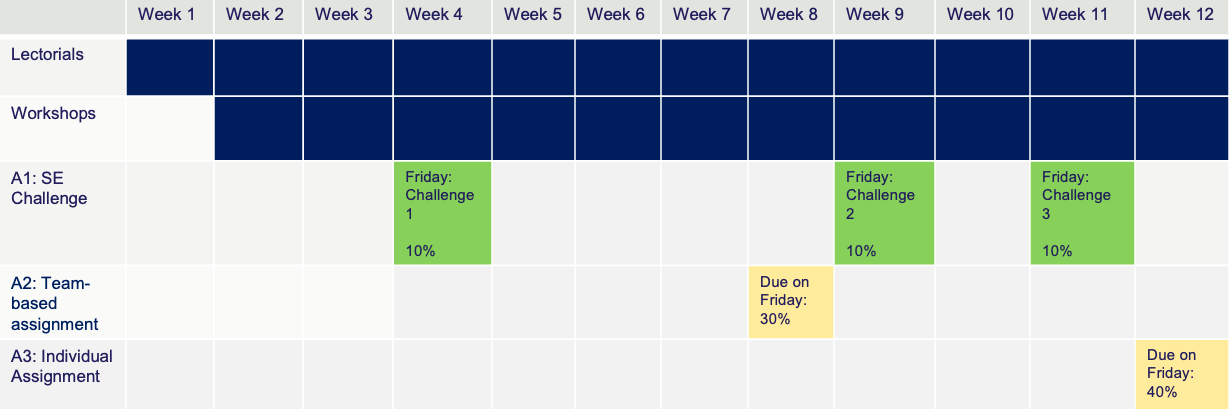}
    \caption{Course structure in 2025} 
     \label{fig:structure2022}
 \end{figure}

 Before 2020, the course delivery was fully face-to-face, and the assessments were designed to have a large in-person exam component, which was 60\% of the final grade: 1h long written mid-semester test in a lecture theatre during a lecture class, 2h long written end-of-semester exam in a large hall. Over 2020-2023, because of the Covid-19 pandemic and correspondingly changing study settings, we adjusted the learning activities in the course as well as our assessment structure. 
 Since 2020, all assessment tasks have been provided in online mode. 
The current course structure is illustrated by Figure~\ref{fig:structure2022}.
The learning activities included in this course are lectorials, workshops, feedback sessions and pre-recorded videos for blended learning, i.e. an integration of classroom face-to-face learning with online learning experiences.

\textbf{Lectorials} are a combination of sessions that have been previously called lectures and tutorials. 
Lectorials are provided each semester week as  2-hour-long sessions over Weeks 1 to 12. In these sessions, we explain key concepts as well as discuss course materials provided as pre-recorded videos, case studies and examples. We also elaborate on key concepts and analyse IT scenarios to identify suitable approaches for project management and requirements engineering activities. These sessions are recorded to support students who were unable to attend the session in person and to provide a basis for student-centred learning.  
To have an interactive recap of the material learned with a lectorial, we include in lectorials additional activities with Mentimeter~\cite{menti}, an online tool that allows engagement with students in real time, as well as interactive quizzes in Canvas, which is the e-learning system adopted in the RMIT university.  
These interactive sessions and quizzes allow students to check whether they have mastered the core concepts and provide a basis for further discussion and clarification if necessary. Mentimeter can also be used to have a student vote on what topics the students would prefer to focus on today, to collect students' input on what issues they would like to discuss, etc. 

   \begin{figure}[ht!]
  \centering  
   {\includegraphics[width=0.31\linewidth]{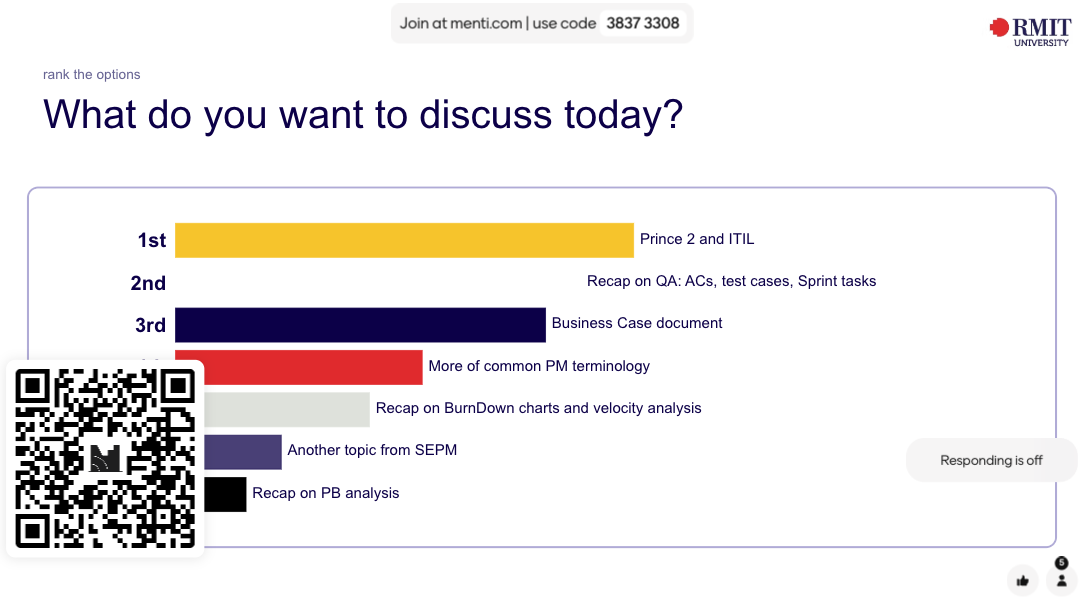}}
  ~~ \includegraphics[width=0.31\linewidth]{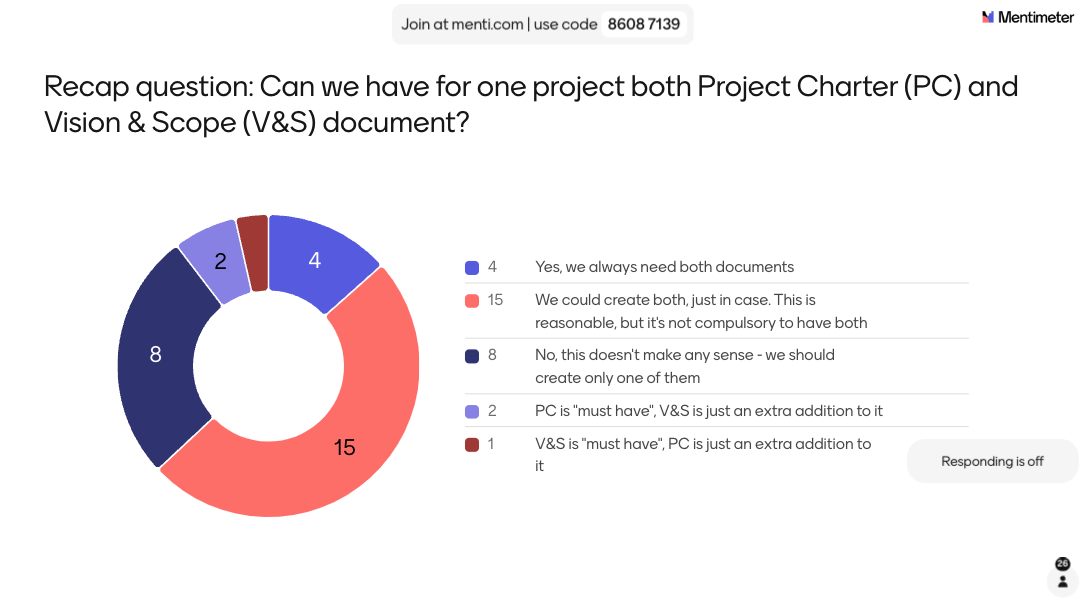} ~~
  \includegraphics[width=0.31\linewidth]{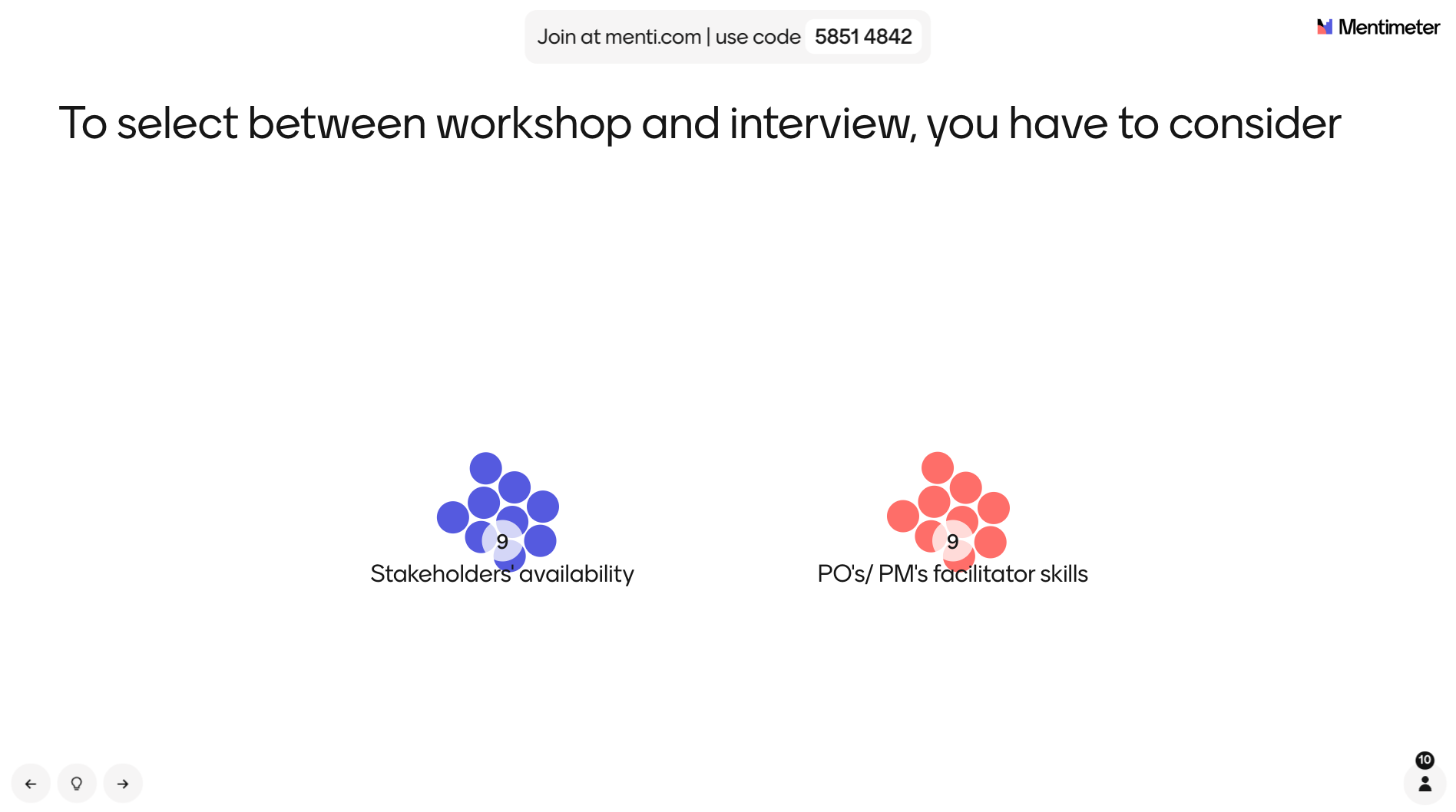}
    \caption{Examples of the application of Mentimeter in the lectoral sessions} 
     \label{fig:menti}
 \end{figure}

\textbf{Workshops} are 2-hour-long sessions provided from Week 2 till Week 12.   
These sessions focus on hands-on experience in working with RE and project management artefacts, using corresponding tools, and providing students with regular feedback on their progress within the assignment tasks.

\textbf{Feedback sessions} are optional drop-in sessions where students can ask questions about the course material. We consider this an important learning activity for private study, where students work on aspects of the course in a self-directed and autonomous manner, allowing for a student-centred approach.  

\textbf{Pre-recorded videos on lecture materials} are provided to support blended learning, which has been a compulsory component in our university over a number of years since the start of the Covid-19 pandemic. The videos cover the core material of the course, which should provide a basis for the lectorial discussions of more in-depth concepts as well as practical case studies. Over 2020-2022, it was expected that students have to watch these videos before attending any lectorial sessions, starting from Week 1. On average, we provided 1 hour of videos per week, with more videos to watch in the first half of the semester, when students have fewer assessment tasks over all courses. However, we observed that this approach is beneficial only for top-performing students, and for students with lower performance the classical approach is easier to follow. As a result, we limited the blended learning components to   the preparation activities on Week 1,
and studying advanced material, which is not compulsory for passing the course.

The assessment structure has also evolved over 2020-2025, to adjust to the changing environment and to support student-centred learning. In what follows, we will discuss the changes and the reasons that led to these changes.

\subsection{Assessment structure: 2021 and 2022}
Over 2021 and 2022, the assessment tasks in the course were divided into two sections: Software Engineering challenge (20\% of the final grade) and a software development project (80\% of the final grade).    

\emph{An end-of-semester SE challenge} was an individual task, that covered such topics as selecting for a given scenario a project management methodology and requirements elicitation techniques, analysing appropriateness of using particular artefacts in a provided scenario, analysing and refining provided traditional requirements and user stories, creating and analysing network diagrams, etc.

\emph{A software development project} has to be conducted using Scrum methodology~\cite{schwaber2011scrum}, with all core artefacts created (product and sprint backlogs, burn-down chart, sprint planning and retro notes). The team size was typically 5-6 students. In both semesters, Bachelor's and Master's students were allocated to different teams, even when the assessment task was the same.  
The teams were provided with 
a real-life scenario and a document imitating unstructured and informal notes made during the initial requirements elicitation interviews with the stakeholders.  The first task for the students was to analyse these provided notes, to identify the details of the functionalities to implement, including their priorities for the project, and to create the corresponding product backlog as a starting point for the Scrum project.  
Thus, all functionalities mentioned in the notes had to be covered in the product backlog, as that would determine the timeline of the project as if it were to happen in real life.

The application to be created was different each semester to avoid plagiarism issues. For example, in Semester 2 of 2021, the project was on the development of an application for real-time updates on the waiting times at the COVID-19 testing sites, while in Semester 1 of 2022 the project was on the development of an application imitating the Wordle game~\cite{wordle}, which became very popular early 2022.   
During the first half of the semester, the students learned the core concepts of PM in general and of Agile and Scrum in particular, including the core PM and RE artefacts.  
In the second half of the semester,  the development sprints start. 
The project-based assessments had the following structure:   
\begin{itemize}
\item  
Part 1 contributes 20\% of the final grade and should be submitted on Week 5 of the semester. By this time, the students should already have mastered the core skill of elaboration and analysis of the Scrum artefacts that provide a basis for the rest of the project: product backlog (PB) and user story cards (USCs). The feedback received from the tutors in Week 4's WPC during the practical workshop session, should be taken into account.   
\item 
Part 2 contributes 42\% of the final grade and has to be submitted on Week 12. %
 This assessment covers the teamwork within the development sprints. 
 Students have to work as a team on a software development project, produce corresponding project management artefacts, as well as analyse their progress and results.
 \item 
 \emph{Work Progress Checks} (WPCs) contributed altogether 18\% of the final grade %
 and should be submitted on Weeks 4, 7, 8, 9, 10, and 11. The aim of this activity was to provide student teams with early feedback on their work. 
 Each WPC was focused on a particular PM/RE/SE artefacts, such as product and sprint backlogs, user story cards, burndown chart, planning and retro notes, etc.%
\end{itemize}
This structure provided authentic assessments and supported the blended learning approach. While the student feedback was overall positive (e.g.,  in 2021, ISYS1106 received the Good Teaching Score (GTS) of 100\% and the Overall Course Satisfaction (OSI) score of 94\%), many students expressed their preference for reducing the proportion of team-based assignments.

\subsection{Assessment structure: 2023-2025}
From 2023, the assessment tasks in the course were divided into three sections: SE challenges, a simulated software development project,  and an individual assessment (on-demand interview).

\emph{(1) SE Challenges (30\% of the final grade):}  
This task comprises 3 challenges, conducted in Weeks 4, 9, and 11. Each challenge will be an individual test that may be taken at any time within 24 hours. Students need to work with practical case studies and corresponding scenarios/artefacts to demonstrate that they can analyse real-life scenarios and conduct SE project management tasks promptly and under time pressure, like in industrial settings. 

\emph{(2) Team-based Assignment on a simulated software development project (30\% of the final grade):} 
This assignment involves elaboration and analysis of the Scrum artefacts that provide a basis for the project: product backlog, user story cards, sprint backlog, meeting notes, etc.  

\emph{(3) Individual Assignment (40\% of the final grade):}  
This assessment is focused on the elaboration and analysis of Scrum artefacts for a provided practical scenario. Students need to work individually to produce corresponding project management artefacts, as well as analyse provided project scenarios and the corresponding artefacts. They also need to demonstrate that you can analyse existing artefacts and provide reasonable feedback on them.

Thus, the following major changes have been made:
 (a) The proportion of individual assessment tasks has been increased in response to the feedback provided by the students. 
 (b)  While having regular WPCs has been welcomed by some students as this provided a more gradual progressive assessment, the majority of students expressed an opinion that this reduced the flexibility of study. Therefore, to provide more student-centred and flexible study settings, the number of submission deadlines has been reduced for the team-based assessment: instead of six WPCs and 2 main submissions, we currently have only one team-based submission in Week 8, while the tasks previously covered by part 2 of the project became an individual task to be submitted in Week 12.   

\section{Findings and lessons learned} 
\label{sec:lessons}
 
In this section, we are going to discuss the lessons learned over 2019-2025 and provide corresponding recommendations. 
The lessons learned are based on our observations during the teaching period, analysis of students' performance over 12 semesters, as well as
analysis of students' feedback provided in the university-run questionnaires.
These questionnaires are run at the end of every semester, and even when they are very general and don't include course-specific questions, they typically uncover some points that might help to improve the course. For example, the proportion of individual assessment tasks has been increased based on the results of these questionnaires.  

\begin{figure}[ht!]
  \centering 
 \includegraphics[width=0.47\linewidth]{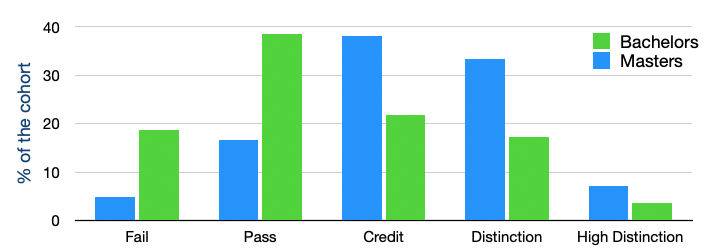}  
    \caption{Normalised grade distribution, Semester 1 2023} 
     \label{fig:grade-distr2023s1}
 \end{figure}

The results of the studies conducted over 2021-24 (in Semesters 1 and 2) demonstrated that overall Master's students performed better than Bachelor's students, which is a general trend over the previous years as well. %
An example of a typical grade distribution is presented in Figure~\ref{fig:grade-distr2023s1}. We haven't observed any significant demographic correlations in terms of student gender. From our observations, the better performance of the Masters' cohort is related to better attention to the deadlines and assessment tasks. 

\subsection{Blended Learning doesn't always work for all students}

Blended learning~\cite{dangwal2017blended,hrastinski2019we,rasheed2020challenges} provides many benefits for highly motivated students, as it allows for more in-depth study of the material during the lectorial time. However, when we deal with large cohorts, not all students are well-motivated. Some students might come unprepared to the class, and this should be taken into account when preparing a plan for a session. Jumping immediately to the advanced concepts, assuming that everyone is already familiar with basic concepts from the pre-lectorial materials, might leave behind the unprepared cohort. 

\emph{Proposed solution:} We propose to have an interactive activity at the beginning of the session to quickly check the level of preparedness of the students, e.g. using anonymous quizzes in Mentimeter~\cite{menti}. The results of interactive quizzes could be immediately discussed with the students, and corresponding materials introduced, if necessary.
Thus, in contrast to the preparation of a ``traditional lecture'', we need to have a very flexible plan on what we discuss with the students in the worst case (when they are totally unprepared), and what we discuss if only some concepts should be covered. %
We also recommend putting only a truly basic/core part of the course material in pre-recorded videos, mostly focusing on theoretical aspects. Otherwise, students might feel overwhelmed by pre-lectural study.

\subsection{Online assessments}

Providing assessments as online, unobserved tasks might increase the risks of cheating. 
Strategies such as supervising an exam remotely, using a Web camera or using \emph{respondus} LockDown Browser to monitor the on-screen actions of students, might disadvantage some students, e.g., if they have a lower speed Internet connection.

\emph{Proposed solutions:} Instead a LockDown Browser approach, we propose to use a combination of project-based tasks and individual timed tasks. 
In the \emph{project-based tasks}, students have to create artefacts for a provided case study and work in a team. 
While grading these tasks, we propose to provide individual grades, calculated based on the team's mark and the individual contributions of each student. 
\emph{Individual timed tasks} are provided as online challenges/quizzes with variants for each question.  
We propose to allow for additional flexibility with a choice of time/date: The task has a limited duration, e.g., one or two hours, and is available for 24h or even for 2-4 days. It's then up to students when to start the task, but as soon as it's started, it must be completed in the allocated time. These settings not only reduce anxiety of having an exam but also reduce the number of requests for a deferred assessment, which we previously had when the exam time overlapped with students' work or family commitments.

   \begin{figure*}[ht!]
  \centering 
 \includegraphics[width=0.65\linewidth]{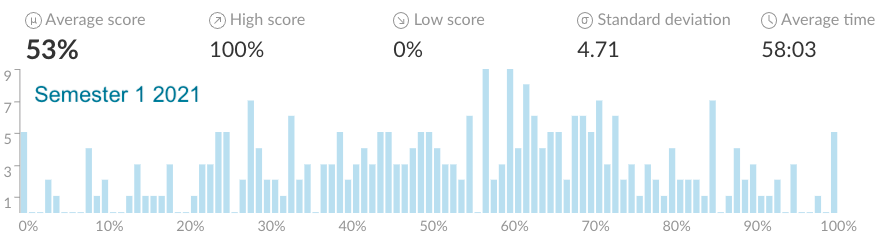}  \\
 ~\\
 \includegraphics[width=0.65\linewidth]{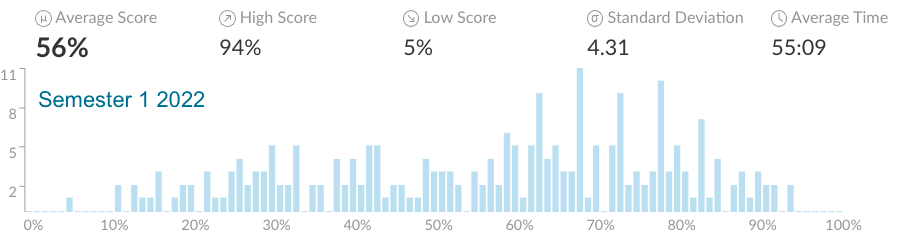}  
    \caption{Performance comparison: Summary for end-of-semester challenges in Semester 1 2021 and Semester 1 2022} 
     \label{fig:test2022}
 \end{figure*}

    \begin{figure*}[ht!]
  \centering  
 \includegraphics[width=0.8\linewidth]{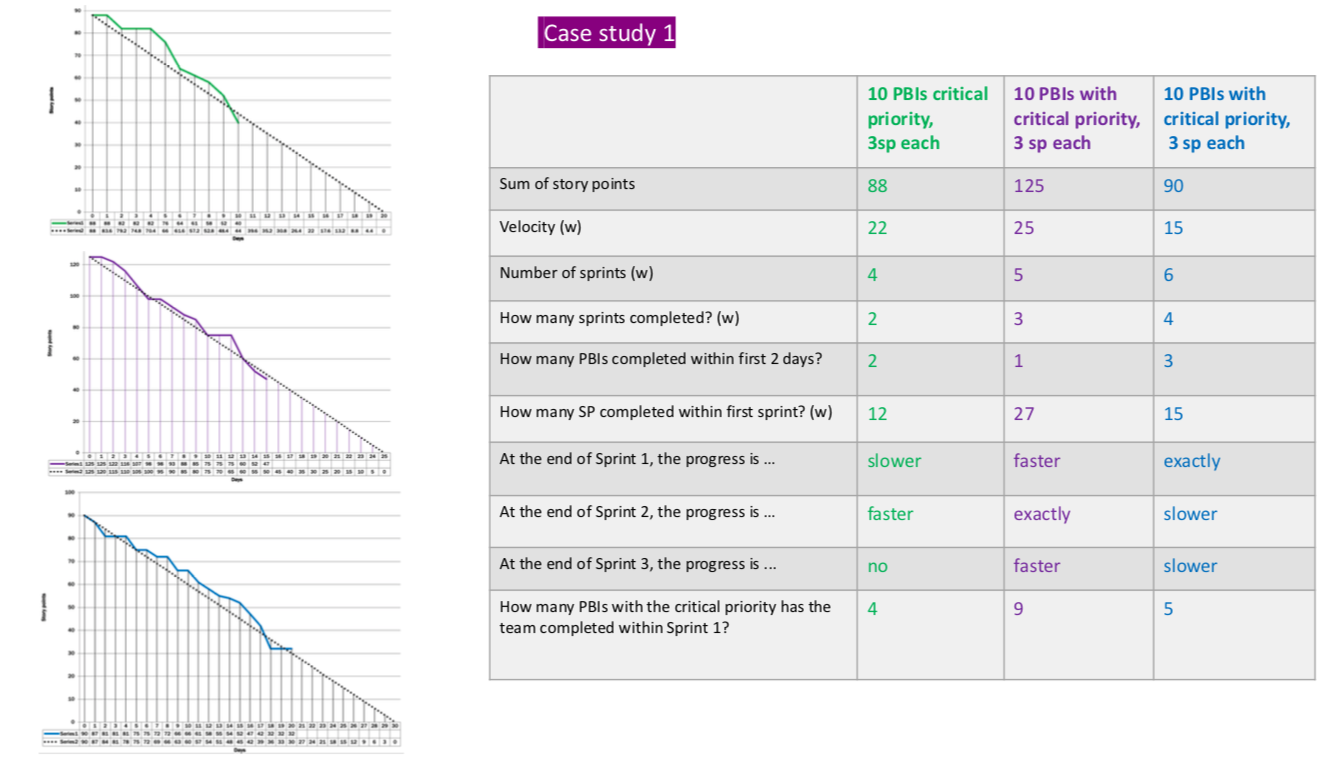}
    \caption{Sample of the marking guide for a challenge case study} 
     \label{fig:question}
 \end{figure*}

In both semesters of 2021, we provided an end-of-semester challenge as a 1h long task with a fixed time/date for everyone, i.e. as it was common for end-of-semester exams. 
From 2022, we have provided challenges as a short (0.5-1h long) task available for 24h, i.e., only the date is fixed. 
Figure~\ref{fig:test2022} presents examples of students' performance in both settings, where x-axis presents the scores achieved by students (from 0\% when all tasks are solved completely incorrectly or none of the tasks has been completed, to  100\%, when all tasks have been completed fully correctly).  The average score (mark) is very similar (53\% in 2021, and 56\% in 2022), so were the average completion times (58:03 and 55:09 minutes respectively), but the standard deviation was larger in 2021 settings: we had some students achieved the maximal score of 100\%, but at the same time some students achieved only 0\%, which both wasn't the case in 2022.   
Also, in 2021 settings we had a larger number of identified cases of cheating, where students submitted answers that matched another variant. To have these cases identifiable, we design the challenge sub-tasks as case studies with several related questions, where a sequence of answers is unique for each variant and cannot be obtained for another variant. Figure~\ref{fig:question} presents a sample marking guide for one of the case studies of the challenge, where the task was to analyse a burn-down chart for a provided scenario.

\subsection{Manual marking using detailed rubrics} 

\begin{figure*}[ht!]
  \centering 
\includegraphics[width=0.5\linewidth]{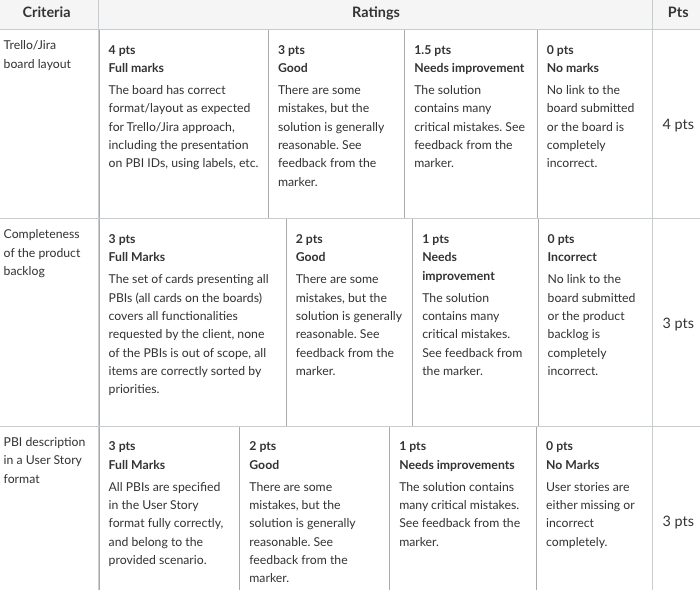}
  \caption{An extract of marking rubrics applied for team-based assessment in 2023-2025} 
\label{fig:rubrics}
\end{figure*}

There is broad evidence that using marking rubrics for assessments might help students support their learning and performance, see, for example, 
\cite{andrade2005student,panadero2013use,reddy2010review,tubino2020authentic}. 
Therefore, we used the rubrics for analysis and marking of students' solutions for all manual marking of the assessment tasks (i.e., marking of all assessments except the SE challenges). Markers had to select one of the ratings associated with each criterion, as well as provide additional remarks to clarify the selection of the rating and what exactly needs to be improved.  To achieve better learning results, the rubrics are shared with students from the beginning of the semester, along with marking rubrics for other assessments within the course. Figure~\ref{fig:rubrics} presents an extract of marking rubrics applied for team-based assessment in 2023-2025.

A lesson we learned regarding using rubrics effectively, is 
that the overall number of points should match the percentage 
the assessment contributes to the final grade. For example, if the assessment is worth 30\% of the final grade, then the rubrics should be presented for 30 points. 
Another lesson learned was that to be effective, rubrics should be based on small, fine-granular criteria. This is especially critical for courses with large cohorts of students, where the teaching team should include many tutors/markers. Having fine-granular criteria allows provide more consistent marking across the whole cohort, which increases student satisfaction and decreases the number of follow-up requests to clarify the provided feedback. Thus,  each criterion of a rubric should be worth max. 5 points, preferably 1-3 points.
 
\subsection{Threats to validity and limitations}
\label{sec:validity}

Several factors might affect the validity of our study. 
Even when we analysed our experience based on relatively large student cohorts over 12 semesters  (2019-2024), the data sample size was limited to cohorts of students from the same university. It would be beneficial to run an extended study with other teaching teams from different universities (ideally, from different countries), with similar course settings.  

Each of the submissions was marked by one member of the teaching team, and the course coordinator double-checked the results of marking only point-wise. In the case some assessments have been marked by the course coordinator, the second marking was provided only for the near-fail cases (where the overall mark for the course is between 44\% and 49\%). It would be good to have a double-marking of all submissions.

It would be also beneficial to conduct interviews with tutors and collect students' perceptions using questionnaires, similar to the approach used by Groher et al. \cite{groher2022exploring}. Currently, we have been using only the results of questionnaires, which are run by the university at the end of each semester.

\section{Related Work} 
\label{sec:background}

There are also a number of works focused on technologies to promote active student learning,  e.g.,~\cite{barry2022virtual,barry2021technologies,ravikumar2024creating,thevathayan2017combining}. 
Many approaches propose using \emph{gamification} of the learning process, e.g.,~\cite{jones2018chirality,yonemura2017practical,spichkova2016boring}. 
Many studies demonstrated that gamification, i.e. \emph{use of game design elements in non-game
contexts}, might enhance learning outcomes by making complex topics more accessible and enjoyable, see, e.g.~\cite{deterding2011game}. 
In our course, we use technology-supported interactive elements as well as gamification elements for recap-quizzes and in-class activities. 
P{\'e}rez-Castillo et al.~\cite{perez2018improving} conducted a study aiming to improve the experience of learning Scrum. The results of their study demonstrated that students get a better learning experience when Scrum concepts are introduced after a comparison with traditional software development methodologies. In our course, we see the comparison as a crucial element (covered by the assessment task  \emph{SE Challenge 3}) because understanding the differences allows students to understand what methodology should be selected in each particular scenario. 
Rodriguez et al. \cite{rodriguez2015virtual} created \emph{Virtual Scrum}, an educational virtual world that simulates a Scrum-based team room. The tool isn't available for practical use in universities, but the idea of visualisation through a virtual environment looks promising.  
Stegh{\"o}fer et al.~\cite{steghofer2017} proposed to use another gamification approach for teaching Scrum: they utilised workshop sessions to build a Lego city in short sprints to focus on the methodological content. While this approach might provide some learning benefits, it is hardly scalable for cohort sizes of 300-500 students. 
Sch{\"a}fer and Burden~\cite{steghofer2022one}  proposed teaching Scrum concepts using the Minecraft game.

In our approach, we use peer feedback to provide individual marks for the team-based assessments. 
There are many studies on the application of peer reviews in education  \cite{gehringer2001electronic,ragone2013peer,sondergaard2012collaborative,topping1998peer}. Most educational approaches to using self/peer feedback see the feedback as an aid for gathering information on the quantity and quality of individual contributions to the team’s work and the individual’s performance as a team member.  For example, Clark et al. \cite{clark2005self}
presented a suite of tools for self and peer feedback and assessment, which was used  in Software Engineering
project for that purpose.   
Anvari et al.  \cite{icse2021-peer-review}  proposed using peer feedback as an alternative method to assessments of artefacts (in the case of their work, conceptual design artefacts). In our previous work~\cite{spichkova2022teaching}, we analysed (1) the level of correlation between students' peer-review skills and their problem-solving skills within Requirements Engineering activities, and
(2) to what extent can the students' peer feedback can be used to provide early feedback within the semester or to mark students' assessments.  
Many studies have demonstrated the successful application of peer reviews within project-based courses for final-year students to enhance learning performance. 
For example, 
Panadero and Alqassab~\cite{panadero2019empirical} presented a study on the importance of anonymity in providing peer feedback. As per the authors, anonymous peer assessment might provide advantages for students’ perceptions about the learning value of peer assessment, as students tend to provide more critical feedback in this setting. 
Garousi \cite{garousi2009applying} incorporates peer reviews in the project-based design project within a final year software engineering course. Evaluation of two offerings of the course demonstrated promising signs of using peer review in the design projects for senior students. 
Richards \cite{richards2009designing} conducted a study with senior students doing a project-based course, and identified that within their course, the peer-review assessment was consistent with staff assessments.

There are many works on AI-assisted marking of student work (auto-grading), but their autonomous application is currently possible only for limited types of tasks, where we can expect the answer in a particular format, as well as following a set of predefined rules. 
For example, closed-ended questions or questions where we expect a numeric answer or an answer consisting of a particular (set of) words, might be auto-graded reliably. While this might sound very restrictive, the tasks of this type might still be authentic. Moreover, they might have a high Bloom's level~\cite{krathwohl2002revision}.  In the current version of our course, we use auto-grading only for marking of SE Challenges, while all other assessments are marked manually. However, the recent advances on Large Language Model (LLM) applications for generating Scrum artefacts and Requirements engineering artefacts indicate that automatisation of these marking tasks might be possible in the coming years, see for example~\cite{arora2024advancing,spichkova2019AI,spichkova2025agile}.

Groher at al. \cite{groher2022exploring} shared their experiences in teaching an introductory programming course to first-year Business Informatics bachelor students, focusing on diversity factors. In the case of our course, diversity aspects might play an even larger role, as the course is currently provided for both Bachelor's and Master's students and has more than 500 enrolments (in Semester 1 2025, the course has 525 enrolments). However, the current study doesn't cover this aspect.

\section{Summary}
\label{sec:summary}

In this paper, we report our experience in designing and teaching a course on SE Project Management for over six years, where the student cohort is very diverse on many levels.  
We aimed to answer the following  research questions: \\
\emph{\textbf{RQ1:}
 How to structure a course to support blended learning? To what extent in what setting a blended learning approach is effective for SE/IT courses?} 
\\
\emph{\textbf{RQ2:} 
What could be a reasonable structure of online assessments, to keep them both authentic and scalable for large cohorts of students?} 

We discussed the lessons learned from applying in our course the following pedagogical methods: (1) blended learning, (2) project-based learning, (3) interactive activities within lectorial sessions, (4) online assessments, and (5) detailed rubrics for manual marking of assessment tasks for large-scale courses with diverse student cohorts.  
We analysed the challenges we observed while teaching the course, and proposed practical and scalable solutions to overcome these challenges. Over the last years, the number of enrolments in SE/CS/IT programs has grown significantly in many universities, therefore, we see the scalability of teaching approaches as one of the core aspects.  

As our future work, we plan to have a more in-depth analysis of the diversity aspects of teaching large-scale cohorts of students. It might also be useful to compare the experiences from several universities/countries.

\bibliographystyle{elsarticle-harv}

\end{document}